\documentclass[12pt]{iopart}

\usepackage{graphicx}
\usepackage[]{subfigure}
\usepackage[Symbol]{upgreek}
\usepackage{citesort}
\begin{document}

\title[Coulomb friction in twisting of biomimetic scale-covered substrate]{Coulomb friction in twisting of biomimetic scale-covered substrate}

\author{Hossein Ebrahimi, Hessein Ali, and Ranajay Ghosh}

\address{Department of Mechanical and Aerospace Engineering, University of Central Florida, Orlando, FL 32816, USA}
\ead{ranajay.ghosh@ucf.edu}
\vspace{0.5pc}
\begin{indented}
\item[]January 2020
\end{indented}

\begin{abstract}
Biomimetic scale-covered substrates provide geometric tailorability via scale orientation, spacing and also interfacial properties of contact in various deformation modes. No work has investigated the effect of friction in twisting deformation of biomimetic scale-covered beams. In this work, we investigate the frictional effects in the biomimetic scale-covered structure by developing an analytical model verified by the finite element simulations. In this model, we consider dry (Coulomb) friction between rigid scales surfaces, and the substrate as the linear elastic rectangular beam. The obtained results show that the friction has a dual contribution on the system by advancing the locking mechanism due to change of mechanism from purely kinematic to interfacial behavior, and stiffening the twist response due to increase the engagement forces. We also discovered, by increasing the coefficient of friction using engineering scale surfaces to a critical coefficient, the system could reach to an instantaneous post-engagement locking. The developed model outlines analytical relationships between geometry, deformation, frictional force and kinematic energy, to design biomimetic scale-covered metamaterials for a wide range of application.
\end{abstract}


\vspace{-1pc}
\submitto{\BB}
%\maketitle

\section{Introduction} \label{Introduction}
Many biological and biomimetic structures use geometrically pronounced features to produce highly nonlinear behavior. These materials include seashells, hierarchical honeycombs, snail spiral, seahorse tail, fish scales, lobster exoskeleton, crab exoskeleton, butterfly wings, armadillo exoskeleton, sponge skeleton, etc. \cite{c07,c08,c09,c10,c11}. Among these structures, dermal scales have garnered special attention recently due to complex mechanical behavior in bending and twisting \cite{c12,c13,c14,c15,c16,c17,c18}. Scales in nature are naturally multifunctional, durable and lightweight \cite{c19,c20,c21,c22,c23,c24,c25,c26,c27,c28}, and protective for the underlying substrate, which has been an inspiration of armor designs \cite{c17,c18,c29,c30} where overlapping scales can resist penetration and provide additional stiffness \cite{c17,c18,c31,c32}. Fabrication methods such as synthetic mesh sewing and stretch-and-release have been recently developed to produce overlapping scale-covered structures in 2D and 1D configuration \cite{c33,c34}. These fabricated structures show almost ten times more puncture resistance than soft elastomers.

However, in addition to these localized loads, global deformation modes such as bending and twisting can be important for a host applications that require a structural mode of deformation such as soft robotics, prosthetics or morphing structures. It is here that characterizing bending and twisting play an important role in ascertaining the benefit of these structures. Prior research has shown that bending and twisting of a substrate show small strain reversible nonlinear stiffening and locking behavior due to the sliding kinematics of the scales in one-dimensional substrates \cite{c35,c36,c37,c38,c39,c40,c41,c42,c43,c44,c48}. The universality of these behavior across bending of uniformly distributed scales, functionally graded scales and uniformly distributed twisting is an important discovery. However, the role of friction and its possible universal role has not been established in literature. In other words, questions remain about the parallels of properties modification brought about by friction in bending with twisting. 

For instance, Coulomb friction in bending regime advances the locking envelopes but at the same time, limits the range of operation \cite{c39}. In the dynamic regime, Coulomb friction can lead to damping behavior, which mimics viscous damping \cite{c44}. Clearly, friction between sliding scales can significantly alter the nature of nonlinearity. However, in spite of these studies, the role of friction in influencing the twisting behavior has never been investigated before.

In this paper we investigate the role of friction in affecting the twisting behavior of biomimetic scale-covered systems under pure torsion for the first time. To this end, we establish an analytical model aided by finite element (FE) computational investigations. We assume rigid scales, linear elastic behavior of the substrate and Coulomb model of friction between scales’ surfaces. We compare our results with FE model to verify the proposed analytical model.

\section{Materials and methods} \label{Materials and methods}

\subsection{Materials and geometry} \label{Materials and geometry}

\begin{figure}[htbp]%
\centering
\subfigure[][]{%
\label{Fig1a}%
\includegraphics[width=3.3in]{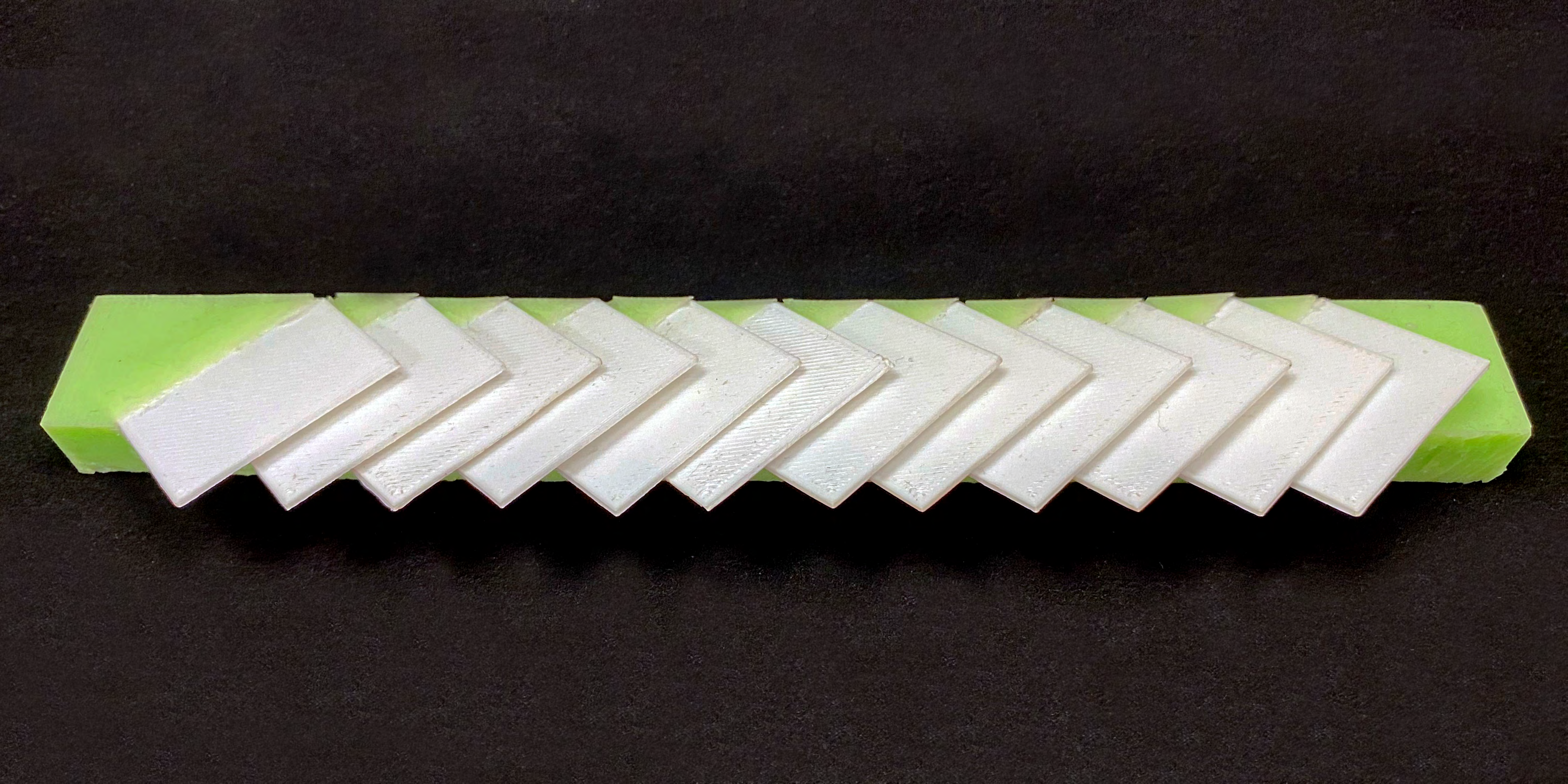}}%
\hspace{8pt}%
\subfigure[][]{%
\label{Fig1b}%
\includegraphics[width=3.3in]{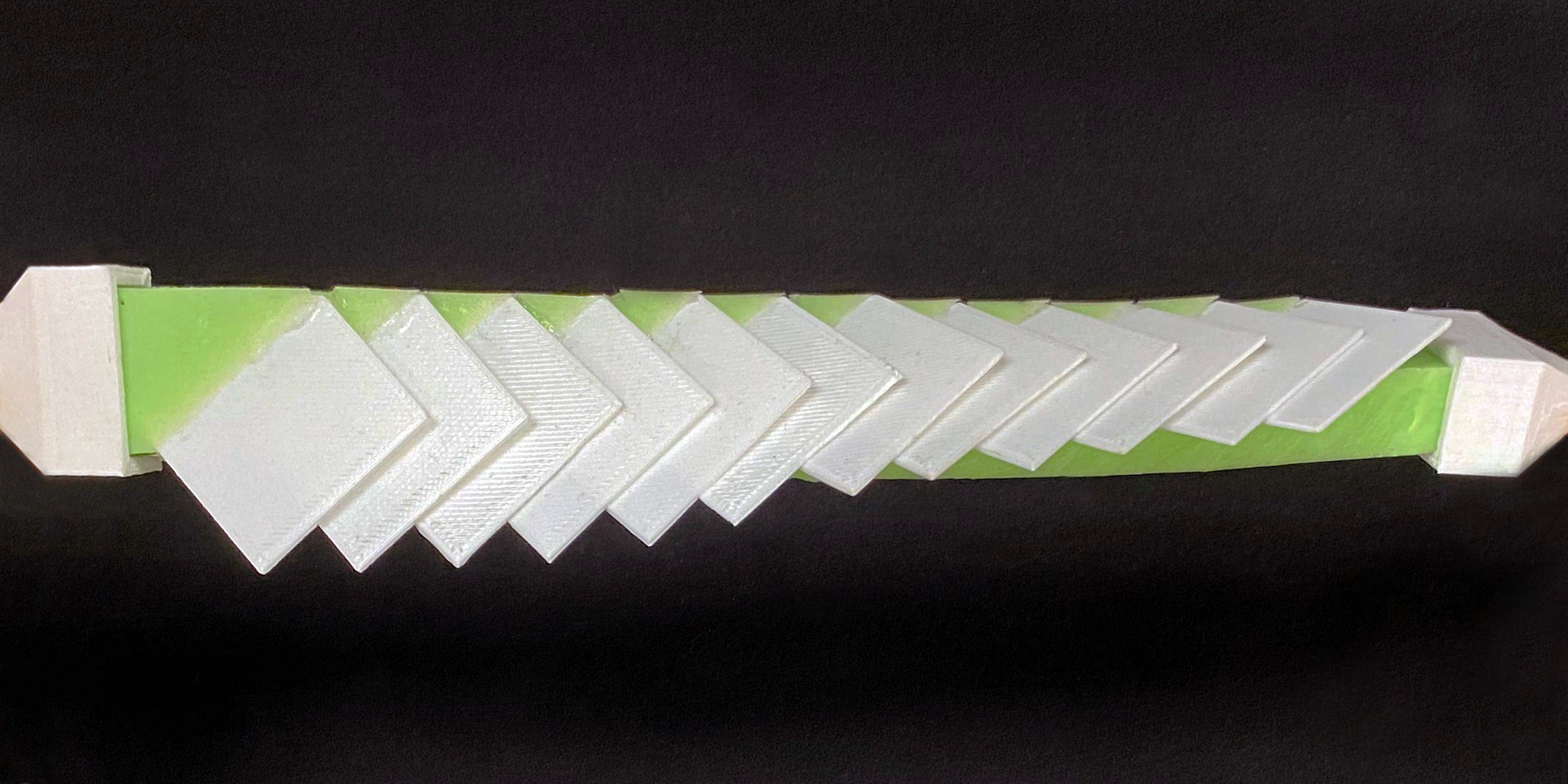}} \\
\caption[]{The fabricated prototype made of 3D-printed PLA scales and molded slender Vinylpolysiloxane (VPS) substrate: 
\subref{Fig1a} untwisted configuration; and
\subref{Fig1b} twisted configuration.}%
\label{Fig1}%
\end{figure}

We consider a rectangular deformable prismatic bar with a row of rigid rectangular plates embedded on substrate’s top surface. For the sake of illustration, we fabricate prototypes of 3D-printed PLA scales ($E_{PLA}\sim3$ $GPa$), embedded onto a silicone substrate and adhered with silicone glue to prefabricated grooves on the molded slender Vinylpolysiloxane (VPS) substrate ($E_{VPS}\sim1.5$ $MPa$) as shown in Figure \ref{Fig1a}. The prototype has been shown under twisting configuration in Figure \ref{Fig1b}. The rigidity assumption is valid in the limit of much higher stiffness of the scales, away from the locking state \cite{c18,c45}.

%%%%%%%%%%%%%%%%%%%% Figure 1 %%%%%%%%%%%%%%%%%%%%
%%%%%%%%%%%%%%%%%%%%%%%%%%%%%%%%%%%%%%%%%%%%%%%%%%

\begin{figure}[htbp]%
\centering
\subfigure[][]{%
\label{Fig2a}%
\includegraphics[width=3.3in]{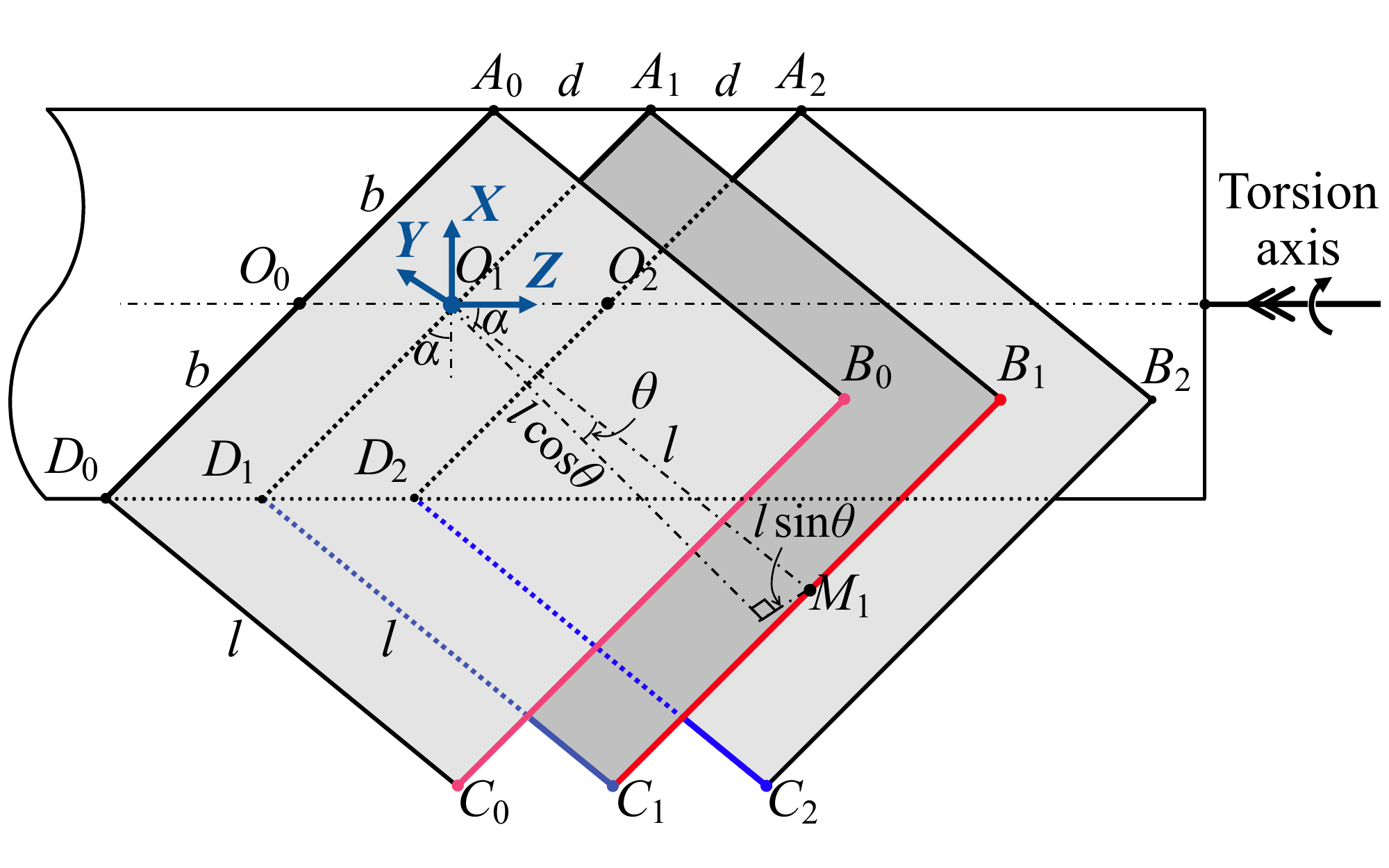}}%
\hspace{8pt}%
\subfigure[][]{%
\label{Fig2b}%
\includegraphics[width=3.3in]{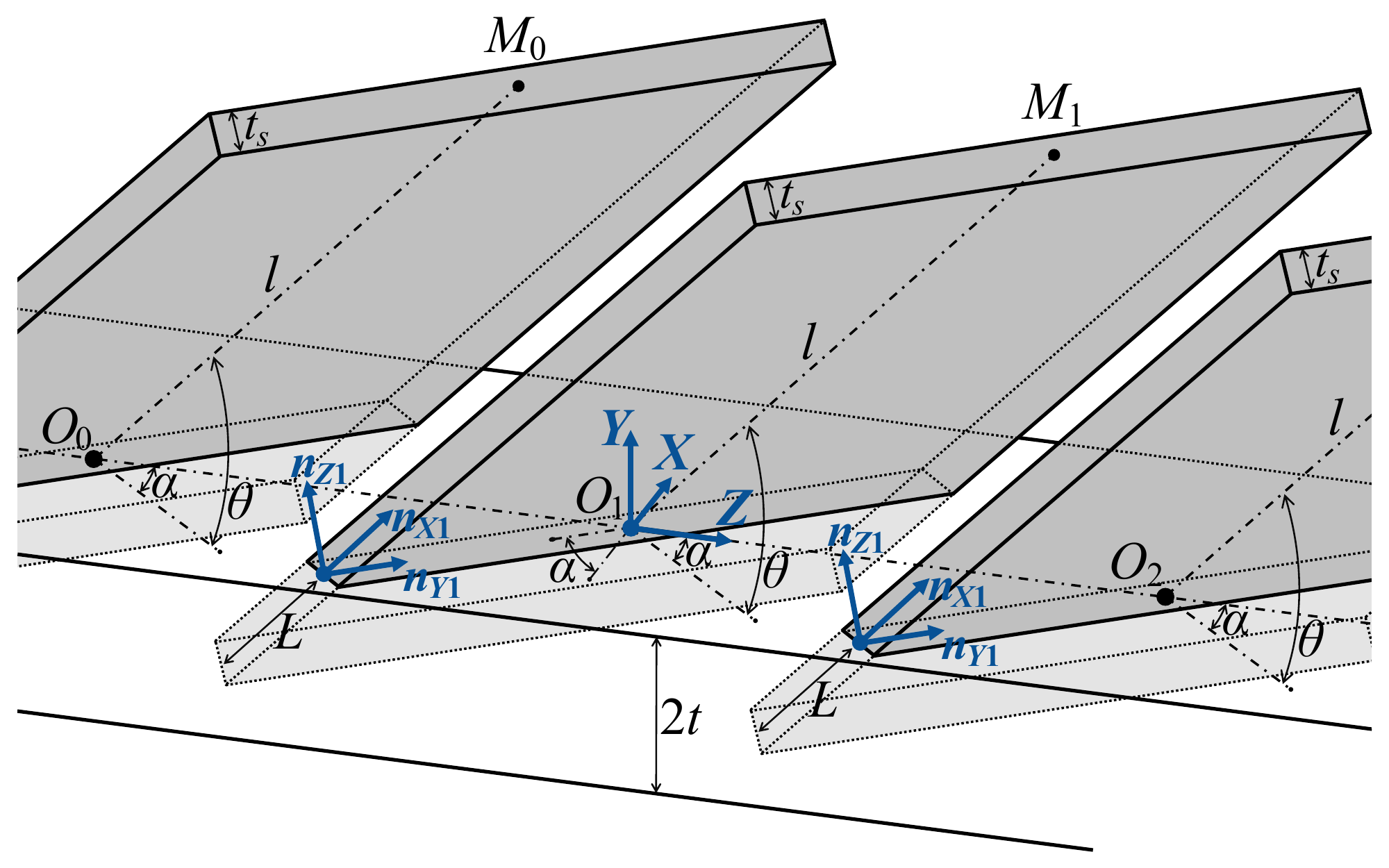}} \\
\caption[]{The schematic of three consecutive scales geometrical configuration: 
\subref{Fig1a} top view of scales configuration; and
\subref{Fig1b} dimetric view to represent scales orientational angles of $\theta$ and $\alpha$, and the embedded part of the each scales. Angle $\theta$ and thickness $t_s$ are exaggerated here.}%
\label{Fig2}%
\end{figure}

The pure twisting behavior allows us to assume periodicity, letting us isolate a fundamental representative volume element (RVE) for modeling the system, Figure \ref{Fig2a}. The scales are considered to be rectangular rigid plates with thickness $t_s$, width $2b$, and length $l_s$, and oriented at angles $\theta$ and $\alpha$ as shown in Figure \ref{Fig2b} with respect to the rectangular prismatic substrate. $\theta$ is the scale inclination angle defined as the dihedral angle between the substrate’s top surface and the scale’s bottom surface, and $\alpha$ is the angle between the substrate’s cross section and the scale’s width. The length of exposed section of scales is denoted as $l$, and the length of embedded section of the scales is $L$. Therefore, the total length of the scale is $l_s=L+l$. The spacing between the scales is constant and denoted by $d$, which is a geometrical parameter reciprocal to the density of scales. We assume that the scale’s thickness $t_s$ is negligible with respect to the length of the scales is $l_s$ ($t_s \ll l_s$), and the scale’s embedded length is also negligible with respect to the substrate’s thickness ($0 \ll L \ll 2t$). This thin-plate idealization for the biomimetic scales is appropriate for this case and typically used in literature for analogous systems \cite{c36,c39,c40,c41,c42,c43,c44}.

%%%%%%%%%%%%%%%%%%%% Figure 2 %%%%%%%%%%%%%%%%%%%%
%%%%%%%%%%%%%%%%%%%%%%%%%%%%%%%%%%%%%%%%%%%%%%%%%%

\subsection{Kinematics} \label{Kinematics}

For global deformation modes such as pure bending and twisting the scale periodicity is a good approximation \cite{c36,c42}. Periodicity assumption allows us to consider just three consecutive scales configuration at the RVE level, We call these scales as ``zeroth scale", ``1\textsuperscript{st} scale", and ``2\textsuperscript{nd} scale" respectively from left to right. Without loss of generality, we consider 1\textsuperscript{st} scale is fixed locally with respect to other scales. A twisting deformation with twist rate $\Upphi$, is applied to the rectangular prismatic substrate about torsion axis, which passes through the beam cross section center. Due to this underlying deformation, the 2\textsuperscript{nd} scale rotates by twist angle of $\varphi=\Upphi d$, and the zeroth scale rotates in reverse direction about the torsion axis with $-\varphi=-\Upphi d$, because 1\textsuperscript{st} scale assumed locally fixed. The continual twisting of the substrate progresses the contact between each two consecutive scales simultaneously due to periodicity, by coincidence between lines $C_1B_1$ and $D_2C_2$, as well as lines $D_1C_1$ and $C_0B_0$. 

To find a contact criterion between 1\textsuperscript{st} scale and 2\textsuperscript{nd} scale, the 3D-equations of lines $C_1B_1$ and $D_2C_2$ would be established. We place the coordinates $XYZ$ on the midpoint of 1\textsuperscript{st} scale's width as shown in Figure \ref{Fig2a}. Then we place coordinates $xyz$ on the torsion axis at point $O=(0,-t,0)$ measured form the coordinates $XYZ$. Hereafter, coordinates $xyz$ is our reference frame. We establish a local coordinates on each scales denoted as ``$i$\textsuperscript{th} scale" and its origin is located on the corner of the scale at point $D_i$. In these local coordinates, the unit vector of $x$-axis ($\bi{{n}_{Xi}}$) is on the edge $D_iC_i$, the unit vector of $y$-axis ($\bi{{n}_{Yi}}$) is on the edge $D_iA_i$, and the unit vector of $z$-axis ($\bi{{n}_{Zi}}$) is out of plane and perpendicular to $\bi{{n}_{Xi}}$ and $\bi{{n}_{Yi}}$, Figure \ref{Fig2b}. On each scales, edges $D_iC_i$ and $A_iB_i$ are parallel and in direction of $\bi{{n}_{Xi}}$, and edges $C_iB_i$ and $D_iA_i$ are parallel and in direction of $\bi{{n}_{Yi}}$. Point $M_i$ is located in the middle of edge $C_iB_i$. Using these established coordinates, symmetric equations of line $C_1B_1$ of 1\textsuperscript{st} scale is as follows \cite{c46}:
%%%%%%%%%%%%%%%%%%%% Equation 1 %%%%%%%%%%%%%%%%
\vspace{0.7pc}
\begin{equation} \label{Eq1}
\frac{x-x_{M_1}}{x_{\bi{{n}_{Y1}}}}=\frac{y-y_{M_1}}{y_{\bi{{n}_{Y1}}}}=\frac{z-z_{M_1}}{z_{\bi{{n}_{Y1}}}},
\end{equation}
\vspace{0.7pc}
%%%%%%%%%%%%%%%%%%%%%%%%%%%%%%%%%%%%%%%%%%%%%%%%%%
where $\bi{{n}_{Y1}}=(x_\bi{{n}_{Y1}},y_\bi{{n}_{Y1}},z_\bi{{n}_{Y1}})$. By putting (\ref{Eq1}) equal to $p$ and using geometrical parameters in Figure \ref{Fig2}, we will have parametric form of the equation of line $C_1B_1$ as follows, where $p$ can vary from $-b$ to $b$:
%%%%%%%%%%%%%%%%%%%% Equation 2 %%%%%%%%%%%%%%%%
\vspace{0.7pc}
\numparts \begin{eqnarray} 
x(p) = p\cos \alpha - l\sin \alpha \cos \theta, \label{Eq2a} \\
y(p) = t + l\sin \theta, \label{Eq2b}\\
z(p) = p\sin \alpha + l\cos \alpha \cos \theta. \label{Eq2c}
\end{eqnarray} \endnumparts %\label{Eq2}
\vspace{-0.7pc}
%%%%%%%%%%%%%%%%%%%%%%%%%%%%%%%%%%%%%%%%%%%%%%%%%%

Point $D_i$ is located at one end of the edge $D_iC_i$. Symmetric equations of line $D_2C_2$ of 2\textsuperscript{nd} scale is as follows:
%%%%%%%%%%%%%%%%%%%% Equation 3 %%%%%%%%%%%%%%%%
\vspace{0.7pc}
\begin{equation} \label{Eq3}
\frac{x-x_{D_2}}{x_{\bi{{n}_{X2}}}}=\frac{y-y_{D_2}}{y_{\bi{{n}_{X2}}}}=\frac{z-z_{D_2}}{z_{\bi{{n}_{X2}}}},
\end{equation}
\vspace{0.7pc}
%%%%%%%%%%%%%%%%%%%%%%%%%%%%%%%%%%%%%%%%%%%%%%%%%%
where $\bi{{n}_{X2}}=(x_\bi{{n}_{X2}},y_\bi{{n}_{X2}},z_\bi{{n}_{X2}})$. To find parametric equation of the line $D_2C_2$, which is on the 2\textsuperscript{nd} scale rotating with angle $\varphi$ about torsion axis, first we locate the corners of 2\textsuperscript{nd} as shown in Figure \ref{Fig2a}, and then their locations are found after rotation, using rotation matrix. Therefore, rotated local coordinates on this scale and the unit vector in direction $D_2C_2$ ($\bi{{n}_{X2}}$) can be established. By using these geometrical parameters and putting (\ref{Eq3}) equal to $q$, we have parametric form of the equation of line $D_2C_2$ as follows, where $q$ can vary from $0$ to $l$:
%%%%%%%%%%%%%%%%%%%% Equation 4 %%%%%%%%%%%%%%%%
\vspace{0.7pc}
\numparts \begin{eqnarray}
x(q) = ( {\tan \theta \tan \varphi - \sin \alpha } )q + ( {t\sin \varphi - b\cos \alpha \cos \varphi }), \label{Eq4a} \\
y(q) = ( {\tan \theta + \sin \alpha \tan \varphi } )q + ( {t\cos \varphi + b\cos \alpha \sin \varphi }), \label{Eq4b} \\ z(q) = ( {\frac{{\cos \alpha }}{{\cos \varphi }}} )q + ( {d - b\sin \alpha } ). \label{Eq4c}
\end{eqnarray} \endnumparts
\vspace{-0.2pc}
%%%%%%%%%%%%%%%%%%%%%%%%%%%%%%%%%%%%%%%%%%%%%%%%%%

To find a contact point between these two lines, (2) and (4) must be identical at $x$, $y$ and $z$ coordinate simultaneously. By putting (\ref{Eq2a}) equal to (\ref{Eq4a}) and also (\ref{Eq2b}) equal to (\ref{Eq4b}) simultaneously, we arrive at the following systems of equations: 
%%%%%%%%%%%%%%%%%%%% Equation 5 %%%%%%%%%%%%%%%%
\vspace{0.7pc}
\begin{equation} \label{Eq5}
\left[\begin{array}{cccc} x_{\bi{{n}_{Y1}}} & -x_{\bi{{n}_{X2}}} \\ y_{\bi{{n}_{Y1}}}& -y_{\bi{{n}_{X2}}}\end{array} \right]  \left[\begin{array}{cccc} p \\ q\end{array} \right] = \left[\begin{array}{cccc} x_{C_2}-x_{M_1} \cr y_{C_2}-y_{M_1}\end{array} \right].
\end{equation}
\vspace{0.7pc}
%%%%%%%%%%%%%%%%%%%%%%%%%%%%%%%%%%%%%%%%%%%%%%%%%%

Solving (\ref{Eq5}) will lead us to equations for $p$ and $q$, and by putting derived equation of $p$ or $q$, into the (\ref{Eq2c}) or (\ref{Eq4c}), yields to an analytical relationship between $\varphi$ and $\theta$. To represent a general form for this relationship, we define dimensionless geometric parameters including $\eta=l/d$, $\beta=b/d$, and $\lambda=t/d$ as the overlap ratio, dimensionless scale width, and dimensionless substrate thickness, respectively. The governing nonlinear relationship between the substrate twist angle $\varphi$ and the scale inclination angle $\theta$ can be written as:
%%%%%%%%%%%%%%%%%%%% Equation 6 %%%%%%%%%%%%%%%%
\vspace{0.7pc}
\begin{eqnarray} \label{Eq6}
\fl (\cos \varphi - 1) {\Big( \beta \sin 2\alpha \sin \theta + \eta {{\cos }^2}\alpha \sin 2\theta + 2\lambda \cos 2\alpha \cos \theta \Big)} - 2\cos \alpha \cos \varphi \sin \theta + \\ 2\sin \alpha \sin \varphi ( {\eta + \lambda \sin \theta } ) + 2\cos \alpha \sin \varphi \cos \theta ( {\beta - \sin \alpha } ) = 0. \nonumber
\end{eqnarray}
\vspace{-0.7pc}
%%%%%%%%%%%%%%%%%%%%%%%%%%%%%%%%%%%%%%%%%%%%%%%%%%

From the beginning of scales engagement, the relationship (\ref{Eq6}) is established between the substrate twist angle $\varphi$ and the scales inclination angle $\theta$. After engaging, scales slide over each other and $\theta$ starts to increase from its initial value $\theta_0$ according to the nonlinear relationship (\ref{Eq6}). Scales engagement start at relatively small twist angle, therefore to find an explicit relationship for the engagement twist angle $\varphi_e$, we linearize (\ref{Eq6}) by considering small twist regime ($\varphi \ll 1$, $\theta \ll 1$) which leads to $\varphi_e={\theta_0}/(\eta \tan \alpha+ \beta- \sin \alpha)$.

Using the kinematic relationship (\ref{Eq6}), we probe the existence of a singular point where locking can take place. This would be the envelope defined by $\partial \varphi /\partial \theta=0$, and beyond which no more sliding is possible without significant deformation of the scales. This point is called the ``kinematic locking" of the system \cite{c42}.

By putting derived equation of $p$ or $q$ into the (2) or (4), we will have the location of point $P_{12}$ as the intersection between lines $D_2C_2$ and $C_1B_1$. We can use the same procedure to establish the locations of zeroth scale's corners and its local coordinates after rotating with angle $-\varphi$ about torsion axis. We find the same nonlinear relationship between $\varphi$ and $\theta$ due to the periodicity of the system, then we can find the location of point $P_{10}$ as the intersection between lines $D_1C_1$ and $C_0B_0$, using the same method.

\subsection{Mechanics} \label{Mechanics}

\begin{figure}[htbp]
\centering
\includegraphics[width=3.3in]{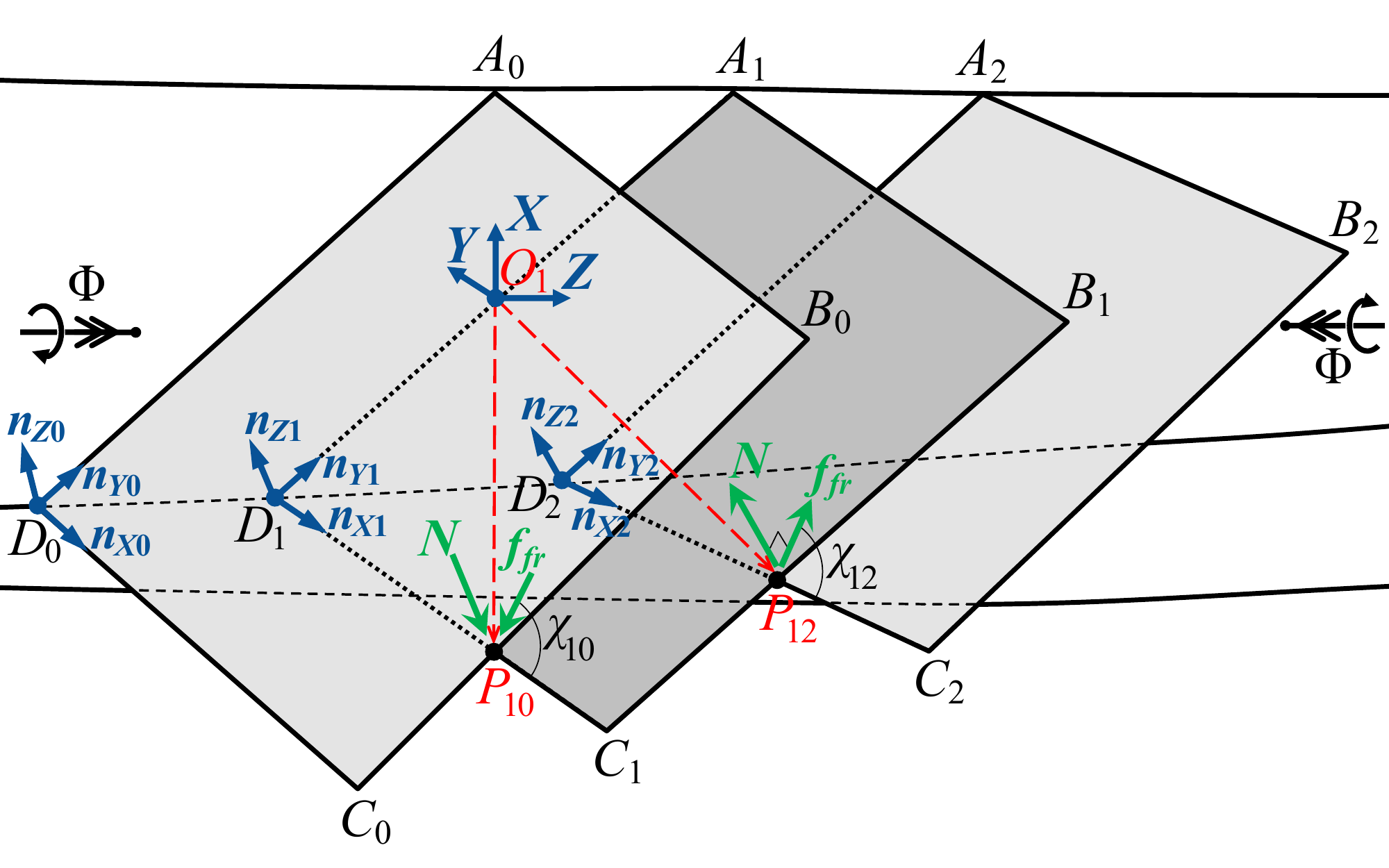}
\caption{Free body diagram of each pair of scales representing their contact points, applied normal force $\bi{N}$, and friction force $\bi{f}_{fr}$ at the contact points.}
\label{Fig3}
\end{figure}

To investigate the role of friction in twisting behavior of biomimetic scale-covered substrate, we investigate the  free body diagram of the RVE (here 1\textsuperscript{st} scale) during engagement as shown in Figure \ref{Fig3}. The forces on the 1\textsuperscript{st} scale are as follows. At contact point between zeroth scale and 1\textsuperscript{st} scale $P_{10}$, there are two reaction forces including friction force $\bi{f_{10}}$ acting in the plane of 1\textsuperscript{st} scale by angle $\chi_{10}$ with respect to the unit vector $\bi{{n}_{X1}}$, and normal force $\bi{N_{10}}$ acting perpendicular to this plane in direction $-\bi{{n}_{Z1}}$ as shown in Figure \ref{Fig3}. Also, at contact point between 1\textsuperscript{st} scale and 2\textsuperscript{nd} scale $P_{12}$, two reaction forces are acting as friction force $\bi{f_{12}}$ in the plane of 2\textsuperscript{nd} scale by angle $\chi_{12}$ with respect to the unit vector $\bi{{n}_{X2}}$, and normal force $\bi{N_{12}}$ perpendicular to the plane of 2\textsuperscript{nd} scale in direction $\bi{{n}_{Z2}}$ as shown in Figure \ref{Fig3}. 

%%%%%%%%%%%%%%%%%%%% Figure 3 %%%%%%%%%%%%%%%%%%%%
%%%%%%%%%%%%%%%%%%%%%%%%%%%%%%%%%%%%%%%%%%%%%%%%%%

Note that the direction of friction forces are dependent on the direction of relative motion between each scale pairs. Due to the periodicity, the value of friction forces are equal $f_{fr}=f_{10}=f_{12}$, and also the value of normal forces are equal $N=N_{10}=N_{12}$. According to the described  free body diagram, the balance of moments at the base of 1\textsuperscript{st} scale can be described in the vectorial format as follows:
%%%%%%%%%%%%%%%%%%%% Equation 7 %%%%%%%%%%%%%%%%
\vspace{0.7pc}
\begin{eqnarray} \label{Eq7}
\fl K_ \theta (\theta - \theta_0)=\bigg( \bi{O_{1}P_{10}} \times \Big( -({f_{fr}} \cos \chi_{10})\bi{{n}_{X1}} - ({f_{fr}} \sin \chi_{10})\bi{{n}_{Y1}} - (N)\bi{{n}_{Z1}} \Big)+ \\
\bi{O_{1}P_{12}} \times \Big( ({f_{fr}} \cos \chi_{12})\bi{{n}_{X2}} + ({f_{fr}} \sin \chi_{12})\bi{{n}_{Y2}} + (N)\bi{{n}_{Z2}} \Big) \bigg).\bi{{n}_{Y1}}, \nonumber
\end{eqnarray}
\vspace{-0.5pc}
%%%%%%%%%%%%%%%%%%%%%%%%%%%%%%%%%%%%%%%%%%%%%%%%%%

\noindent where $\bi{O_{1}P_{10}}$ and $\bi{O_{1}P_{12}}$ are the position vector of contact points $P_{10}$ and $P_{12}$ with respect to the base of the 1\textsuperscript{st} scale, respectively as shown in Figure \ref{Fig3}. $K_{\theta}$ is the ``rotational spring constant" or the ``rigid scale--elastic substrate joint stiffness". As the scales engage, they tend to push each other and increase their inclination angle $\theta$, but the elastic substrate resists against scales rotation. This resistance is modeled as linear torsional spring \cite{c35,c36}, and the absorbed energy due to the rotation of each scale is $U_{scale}=\frac{1}{2} {K}_\theta (\theta-\theta_0 )^2$, thus the local reaction moment would be $M_{scale}=K_\theta (\theta-\theta_0)$. According to developed scaling expression in \cite{c42}, $K_{\theta}=3.62{E_B}{t_s}^2b( {{L}/{t_s}})^{1.55}$, where $E_B$ is the elastic modulus of substrate.

To describe the relative motion between zeroth scale and 1\textsuperscript{st} scale, we would need the relative motion of contact point $P_{10}$ on the edge $D_1C_1$ and edge $C_0B_0$. Motion of point $P_{10}$ on the edge $D_1C_1$ can be described as the change in the length of vector $\bi{P_{10}C_1}$, which is always in direction of $\bi{{n}_{X1}}$, and the change in the length of vector $\bi{P_{10}C_0}$, which is always in direction of $\bi{{n}_{Y0}}$. By using the superposition principle, the total differential displacement of point $P_{10}$ can be described in vectorial format as $\mathrm{d}\bi{R_{10}}=\big(\mathrm{d}|\bi{P_{10}C_1}|\big)\bi{{n}_{X1}}+\big(\mathrm{d}|\bi{P_{10}C_0}|\big)\bi{{n}_{Y0}}$, Figure \ref{Fig3}. The unit vector $\bi{{n}_{Y0}}$ can be described in the local coordinate established on 1\textsuperscript{st} scale as follows:
%%%%%%%%%%%%%%%%%%%% Equation 8 %%%%%%%%%%%%%%%
\vspace{0.7pc}
\begin{equation} \label{Eq8}
\bi{{n}_{Y0}}=(\bi{{n}_{Y0}}.\bi{{n}_{X1}})\bi{{n}_{X1}}+(\bi{{n}_{Y0}}.\bi{{n}_{Y1}})\bi{{n}_{Y1}}+(\bi{{n}_{Y0}}.\bi{{n}_{Z1}})\bi{{n}_{Z1}}.
\end{equation}
\vspace{0.7pc}
%%%%%%%%%%%%%%%%%%%%%%%%%%%%%%%%%%%%%%%%%%%%%%%%%%

By projecting $\bi{{n}_{Y0}}$ on the 1\textsuperscript{st} scale plane, we can describe relative motion of zeroth scale with respect to 1\textsuperscript{st} scale as the planar relative displacement, as follows: 
%%%%%%%%%%%%%%%%%%%% Equation 9 %%%%%%%%%%%%%%%
\vspace{0.7pc}
\begin{equation} \label{Eq9}
\fl \mathrm{d}\bi{r}=\Big(\mathrm{d}|\bi{P_{10}C_1}|+\mathrm{d}|\bi{P_{10}C_0}|(\bi{{n}_{Y0}}.\bi{{n}_{X1}})\Big)\bi{{n}_{X1}}+ \Big(\mathrm{d}|\bi{P_{10}C_0}|(\bi{{n}_{Y0}}.\bi{{n}_{Y1}})\Big)\bi{{n}_{Y1}}.
\end{equation}
\vspace{0.7pc}
%%%%%%%%%%%%%%%%%%%%%%%%%%%%%%%%%%%%%%%%%%%%%%%%%%

The length of (\ref{Eq9}) can be described as the relative differential displacement value: 
%%%%%%%%%%%%%%%%%%%% Equation 10 %%%%%%%%%%%%%%%
\vspace{0.7pc}
\begin{equation} \label{Eq10}
\fl \mathrm{d}r=|\mathrm{d}\bi{r}|=\sqrt{\Big(\mathrm{d}|\bi{P_{10}C_1}|+\mathrm{d}|\bi{P_{10}C_0}|(\bi{{n}}_{Y0}.\bi{{n}_{X1}})\Big)^2+ \Big(\mathrm{d}|\bi{P_{10}C_0}|(\bi{{n}_{Y0}}.\bi{{n}_{Y1}})\Big)^2}.
\end{equation}
\vspace{0.7pc}
%%%%%%%%%%%%%%%%%%%%%%%%%%%%%%%%%%%%%%%%%%%%%%%%%%

To find the angle between the friction force $\bi{{f}_{fr}}$ acting in the plane of 1\textsuperscript{st} scale and the unit vector $\bi{{n}_{X1}}$, we can use (\ref{Eq9}) and (\ref{Eq10}) as the relative displacement vector and its value, then angle $\chi_{10}$ is derived as:
%%%%%%%%%%%%%%%%%%%% Equation 11 %%%%%%%%%%%%%%%
\vspace{0.7pc}
\begin{equation} \label{Eq11}
\chi_{10} = \arccos \Big( \frac{1}{\mathrm{d}r} \big( \mathrm{d}|\bi{P_{10}C_1}|+\mathrm{d}|\bi{P_{10}C_0}|(\bi{{n}_{Y0}}.\bi{{n}_{X1}}) \big) \Big).
\end{equation}
\vspace{0.7pc}
%%%%%%%%%%%%%%%%%%%%%%%%%%%%%%%%%%%%%%%%%%%%%%%%%%

If we repeat similar steps for the relative motion between 1\textsuperscript{st} scale and 2\textsuperscript{nd} scale, it will lead to the similar relationship for the angle between the friction force $\bi{{f}_{fr}}$ acting in the plane of 2\textsuperscript{nd} scale and the unit vector $\bi{{n}_{X2}}$. Finally by computing the values of these relationships, we find that $\chi_{10} = \chi_{12}$, and can be shown as ${\chi}$. This finding also conform the periodicity in the system.

According to the Coulomb's Law of Friction, scales do not slide while $f_{fr} \leq \mu N$, where $\mu$ and $N$ are coefficient of friction and normal force, respectively, while sliding regime is marked by the equality. Note that we use the same value for static coefficient of friction as well as the kinetic coefficient of friction in this study, although typically static coefficient of friction is slightly higher. Using these considerations, we can derive the following expression as the non-dimensionalized friction force $\overline{f}_0$, with respect to the free body diagram shown in Figure \ref{Fig3}:
%%%%%%%%%%%%%%%%%%%% Equation 12 %%%%%%%%%%%%%%%
\vspace{0.7pc}
\begin{eqnarray} \label{Eq12}
\fl \overline{f}_0 = \frac{f_{fr} l}{K_\theta} \leq \\
\fl \frac{(\theta - \theta_0)l}{
 \Big( \hspace{-2pt} \bi{O_{1}P_{12}} \hspace{-3pt} \times \hspace{-3pt} \big( \hspace{-2pt} \cos \chi \bi{{n}_{X2}} \hspace{-2pt} + \hspace{-2pt} \sin \chi \bi{{n}_{Y2}} \hspace{-2pt} + \hspace{-2pt} \frac{\bi{{n}_{Z2}}}{\mu} \hspace{-2pt} \big) \hspace{-2pt} - \hspace{-2pt}
 \bi{O_{1}P_{10}} \hspace{-3pt} \times \hspace{-3pt} \big( \hspace{-1pt} \cos \chi \bi{{n}_{X1}} \hspace{-2pt} + \hspace{-2pt} \sin\chi \bi{{n}_{Y1}} \hspace{-2pt} + \hspace{-2pt} \frac{\bi{{n}_{Z1}}}{\mu} \hspace{-1pt} \big) \hspace{-1pt} \Big).\bi{{n}_{Y1}}}. \nonumber
\end{eqnarray}
\vspace{0.1pc}
%%%%%%%%%%%%%%%%%%%%%%%%%%%%%%%%%%%%%%%%%%%%%%%%%%

Due to the nature and the geometrical configuration of the system, the magnitude of the friction force derived in (\ref{Eq12}), may exhibit singularity at a certain twist rate. This rise in friction force may lead to a ``frictional locking" mechanism, observed in the bending case \cite{c39}. If predicted, the frictional locking should happen at the lower twist rate rather than the kinematic locking, because of the limiting nature of friction force. We call the twist rate in which locking happens as $\Upphi_{lock}$, and the twist angle and the scale inclination angle would be as $\varphi_{lock}=\Upphi_{lock}d$ and $\theta_{lock}$, respectively.

The friction force computed above will lead to dissipative work in the system during sliding. The non-dissipative component of the deformation is absorbed as the elastic energy of the biomimetic beam. This elastic energy is composed of elastic energy of the beam and the scales rotation. To calculate this elastic energy of the beam, we consider a linear elastic behavior for the beam with a warping coefficient $C_w$ for a non-circular beam \cite{c42,c47}. Furthermore, due to the finite embedding of the scales, there will be an intrinsic stiffening of the structure even before scales engagement. This stiffening can be accurately captured by using an inclusion correction factor $C_f$ \cite{c42}. $C_f$ is function of the volume fraction of the rigid inclusion into the elastic substrate, and postulated as $C_f=1+1.33({\zeta \beta}/{\lambda})$, where $\zeta=L/d$ for an analogous system \cite{c42}. With these considerations, modified torque-twist relationship of the beam is $T=C_f C_w G_B I\Upphi$, and the elastic energy of the beam can be considered as $U_B=\frac{1}{2} C_f C_w G_B I\Upphi^2$. As mentioned earlier, the energy absorbed by the scales can be obtained by assuming the scale’s resistance as linear torsional spring and the absorbed energy due to the rotation of each scale will be $U_{scale}=\frac{1}{2} K_\theta (\theta-\theta_0)^2$. Similarly the dissipation can be given as the product of the sliding friction and distance travelled by the point of application per scale. Then we use the work–energy balance to arrive at:
%%%%%%%%%%%%%%%%%%%% Equation 13 %%%%%%%%%%%%%%%
\vspace{0.7pc}
\begin{equation} \label{Eq13}
\fl \int_{0}^{\Upphi} T( {\Upphi }')\mathrm{d}{\Upphi}' = {\frac{1}{2}}{C_f}{C_w}{G_B}I{{\Upphi}^2} + \bigg( {\frac{1}{2}} {\frac{1}{d}}{K_\theta}{( {\theta - {\theta _0}})^2} + {\frac{1}{d}} \int_{\Upphi_e}^{\Upphi} \hspace{-3pt} f_{fr} \mathrm{d}r \bigg) H(\mathrm{\Upphi} - {{\Upphi}_e} ), 
\end{equation}
\vspace{0.7pc}
%%%%%%%%%%%%%%%%%%%%%%%%%%%%%%%%%%%%%%%%%%%%%%%%%%
where $\Upphi$, $\Upphi_e=\varphi_e/d$, $G_B$, and $I$ are the current twist rate, the engagement twist rate, the shear modulus of elasticity, and the cross section's moment of inertia of the beam. $H(\Upphi-\Upphi_e)$ is the Heaviside step function to track scales engagement. Also, $C_f$, $C_w$, and $K_\theta$ are inclusion correction factor, warping coefficient, and rotational spring constant of scale–substrate joint stiffness, respectively. In (\ref{Eq13}), $f_{fr}$ is representing the friction force between scales, and $\mathrm{d}r$ is the relative differential displacement described in (\ref{Eq10}).

The torque--twist rate relationship for the substrate's unit length could be obtained by taking the derivative of (\ref{Eq13}) with respect to the twist rate $\Upphi$, while considering $\varphi=\Upphi d$, as follows:
%%%%%%%%%%%%%%%%%%%% Equation 14 %%%%%%%%%%%%%%%
\vspace{0.7pc}
\begin{equation} \label{Eq14}
 T( \Upphi)={C_f}{C_w}{G_B}I{\Upphi} + \bigg({K_\theta}(\theta - \theta _0)\frac{\partial\theta}{\partial\varphi} + f_{fr} \frac{\mathrm{d}r}{\mathrm{d}\varphi} \bigg) H(\mathrm{\Upphi} - {{\Upphi}_e}). 
\end{equation}
\vspace{0.7pc}
%%%%%%%%%%%%%%%%%%%%%%%%%%%%%%%%%%%%%%%%%%%%%%%%%%

We also compute the maximum possible dissipation of the system by computing the frictional work done till locking ($W_{fr}$) and compare it with the total work done ($W_{sys}=U_{el}+W_{fr}$, where $U_{el}$ is the elastic energy of the system). These energies can be computed per unit length of the beam as:
%%%%%%%%%%%%%%%%%%%% Equation 15 %%%%%%%%%%%%%%%%
\vspace{0.7pc}
\numparts \begin{eqnarray}
U_{el}={\frac{1}{2}}\bigg({C_f}{C_w}{G_B}I{(\Upphi_{lock})^2} +{\frac{1}{d}}{K_\theta}{( {\theta_{lock} - {\theta _0}})^2 \bigg)}, \label{Eq15a} \\
W_{fr}={\frac{1}{d}} \int_{\Upphi_e}^{\Upphi_{lock}} \hspace{-3pt} f_{fr} \mathrm{d}r. \label{Eq15b}
\end{eqnarray} \endnumparts
\vspace{-0.2pc}
%%%%%%%%%%%%%%%%%%%%%%%%%%%%%%%%%%%%%%%%%%%%%%%%%%

We define the relative energy dissipation ($RED$) factor as the ratio of the frictional work per unit length $W_{fr}$, to the total work done on the system per unit length $W_{sys}$:
%%%%%%%%%%%%%%%%%%%% Equation 16 %%%%%%%%%%%%%%%
\vspace{0.7pc}
\begin{equation} \label{Eq16}
RED=\frac{W_{fr}}{W_{sys}}. 
\end{equation}
\vspace{0.7pc}
%%%%%%%%%%%%%%%%%%%%%%%%%%%%%%%%%%%%%%%%%%%%%%%%%%

Generally, $RED$ is dependant on the coefficient of friction $\mu$, dimensionless geometric parameters of the system $\eta$, $\beta$, and $\lambda$, scale spacing $d$, scales initial orientation angles $\alpha$ and $\theta_0$, substrate elastic properties $G_B$, $I$, and $C_w$, and scale--substrate joint parameters $K_{\theta}$ and $C_f$, but the most important parameters are $\mu$, $\eta$, and $\alpha$.

\section{Finite element simulations} \label{Finite element simulations}

We developed an FE model for verification of the analytical model of the biomimetic scale-covered system under twisting deformation. The FE simulations are carried out using commercially available software ABAQUS/CAE 2017 (Dassault Syst\`emes). We considered 3D deformable solids for scale and substrate. However, for the scales, rigid body constraint was imposed. A sufficient substrate length is considered for rectangular prismatic substrate to satisfy the periodicity. Then an assembly of substrate with a row of 25 scales embedded on its top surface is created. The scales are oriented at angles of $\theta_0$ and $\alpha$ as defined in the analytical model. Linear elastic material properties including $E_B$ and $\nu$ are applied to the substrate part which leads to the shear modulus of $G_B={\frac{E_B}{2(1+\nu)}}$. 

The simulation considered as a static step with nonlinear geometry option. The left side of the beam is fixed and the twisting load was applied on the other side of the beam. A frictional contact criteria is applied to the scales surfaces with coefficient of friction $\mu$ for a twisting simulation. The top layer of substrate is meshed with tetrahedral quadratic elements C3D10 due to the geometrical complexity around scales inclusion. Quadratic hexahedral elements C3D20 are used for other regions of the model. A mesh convergence study is carried out to find sufficient mesh density for different regions of the model. A total of almost 70,000 elements are employed in the FE model.

\section{Results and discussion} \label{Results and discussion}

\begin{figure}[htbp]
\centering
\includegraphics[width=3in]{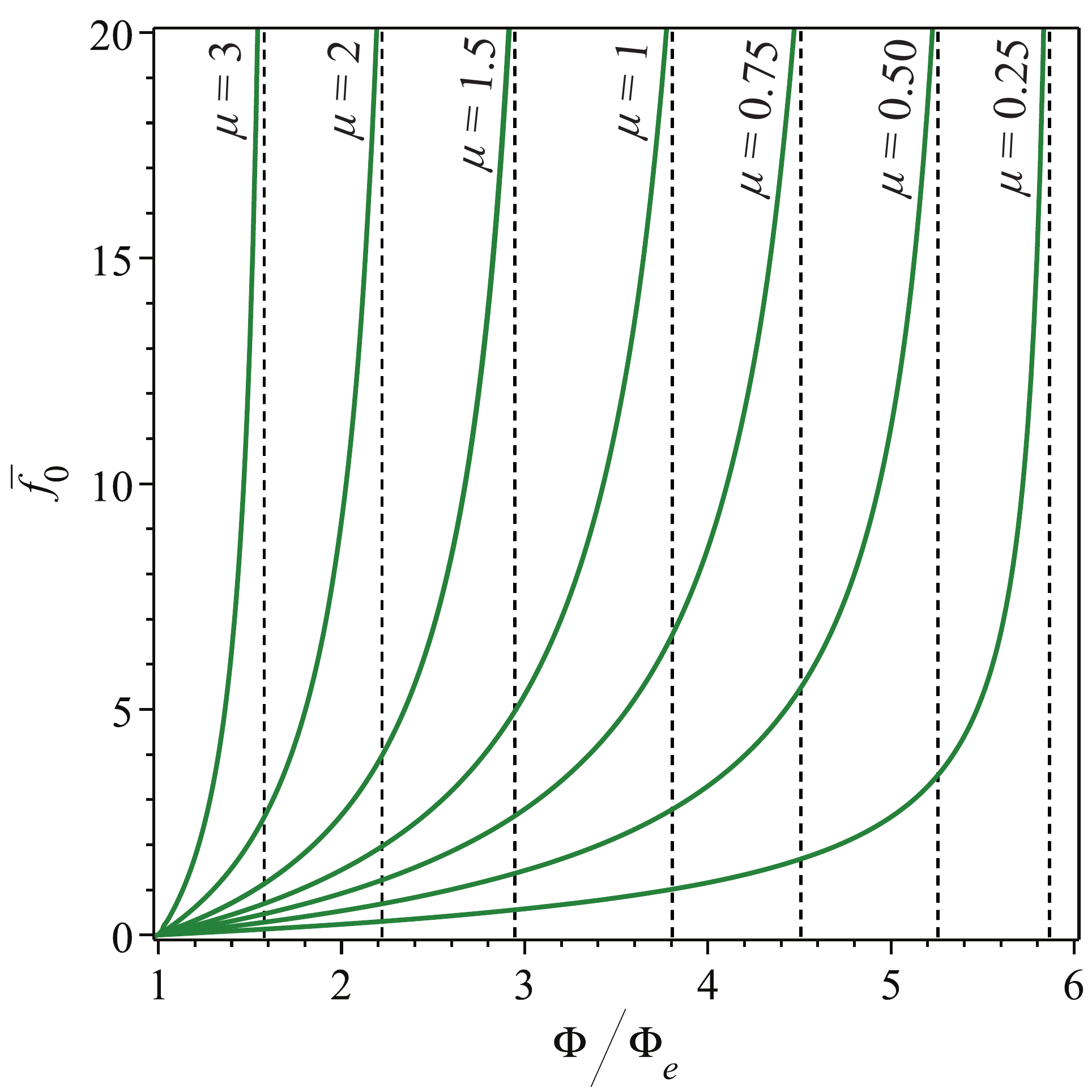}
\caption{Non-dimensionalized friction force vs Non-dimensionalized twist rate ($\Upphi_e$ is the engagement twist rate) for various coefficients of friction with the given values of $\eta=3$, $\theta_0=10^\circ$, $\alpha=45^\circ$, $\beta=1.25$, and $\lambda=0.45$. This figure shows that the friction forces approach singularity near a certain twist rate as the frictional locking configuration for each $\mu$.}
\label{Fig4}
\end{figure}

To study the frictional force behavior in this system, we use (\ref{Eq12}) to plot non-dimensionalized friction force $\overline{f}_0$ for different $\mu$ values at various non-dimensionalized twist rate $\Upphi/\Upphi_e$. This is shown in Figure \ref{Fig4} for a system with $\eta=3$, $\theta_0=10^\circ$, $\alpha=45^\circ$, $\beta=1.25$, and $\lambda=0.45$. From this figure, it is clear that increasing twist leads to a rapid increase in the friction force for any coefficient of friction. There is a singular characteristic to this load as shown with dashed lines for each $\mu$ in Figure \ref{Fig4}, which indicates a friction based locking mechanism. This is in addition to the purely kinematic locking mechanism reported earlier in literature for frictionless counterparts \cite{c42}. We call the value of twist rate at the locking point, as the locking twist rate $\Upphi_{lock}$. 

%%%%%%%%%%%%%%%%%%%% Figure 4 %%%%%%%%%%%%%%%%%%%%
%%%%%%%%%%%%%%%%%%%%%%%%%%%%%%%%%%%%%%%%%%%%%%%%%%

\begin{figure}[htbp]%
\centering
\subfigure[][]{%
\label{Fig5a}%
\includegraphics[width=3in]{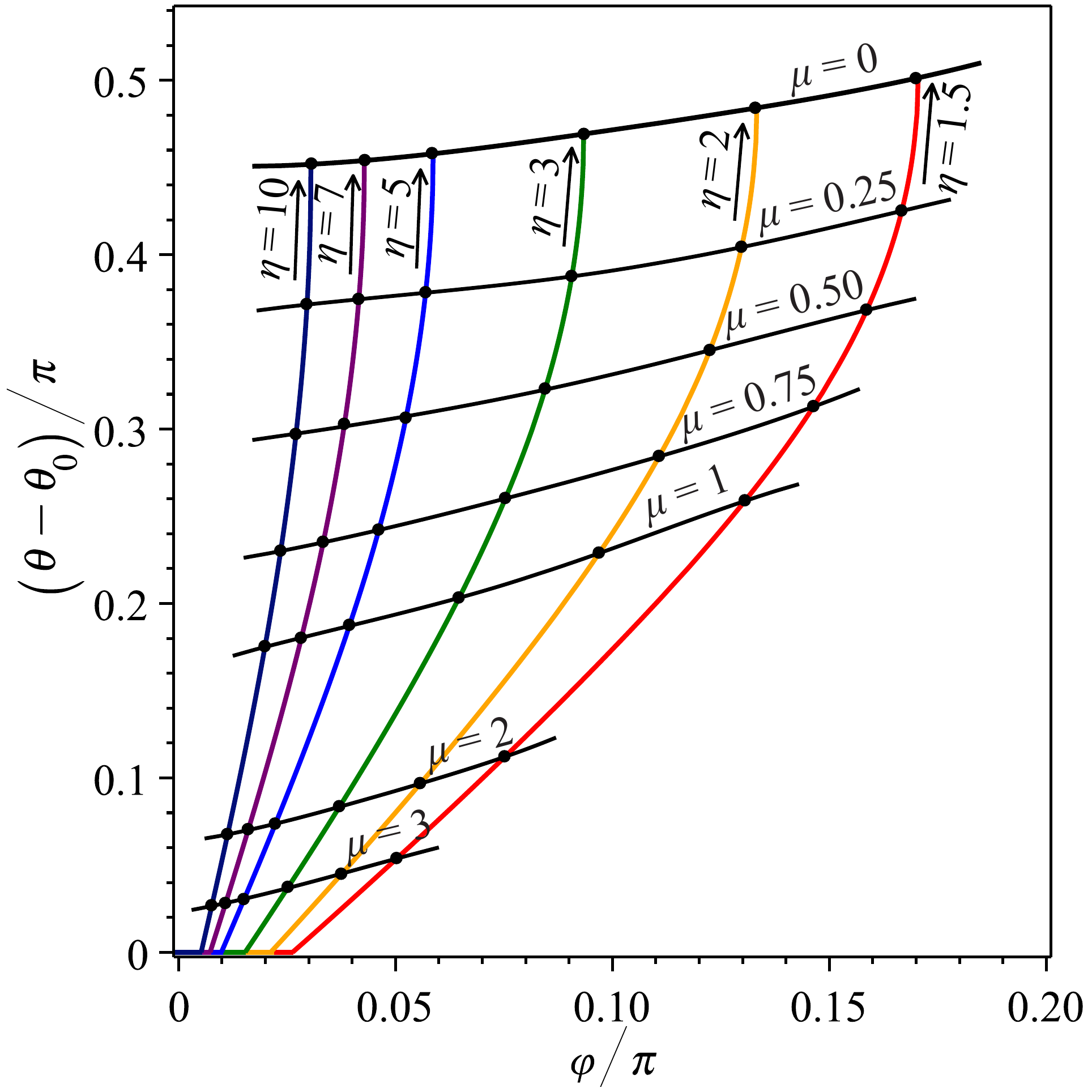}}%
\hspace{8pt}%
\subfigure[][]{%
\label{Fig5b}%
\includegraphics[width=3in]{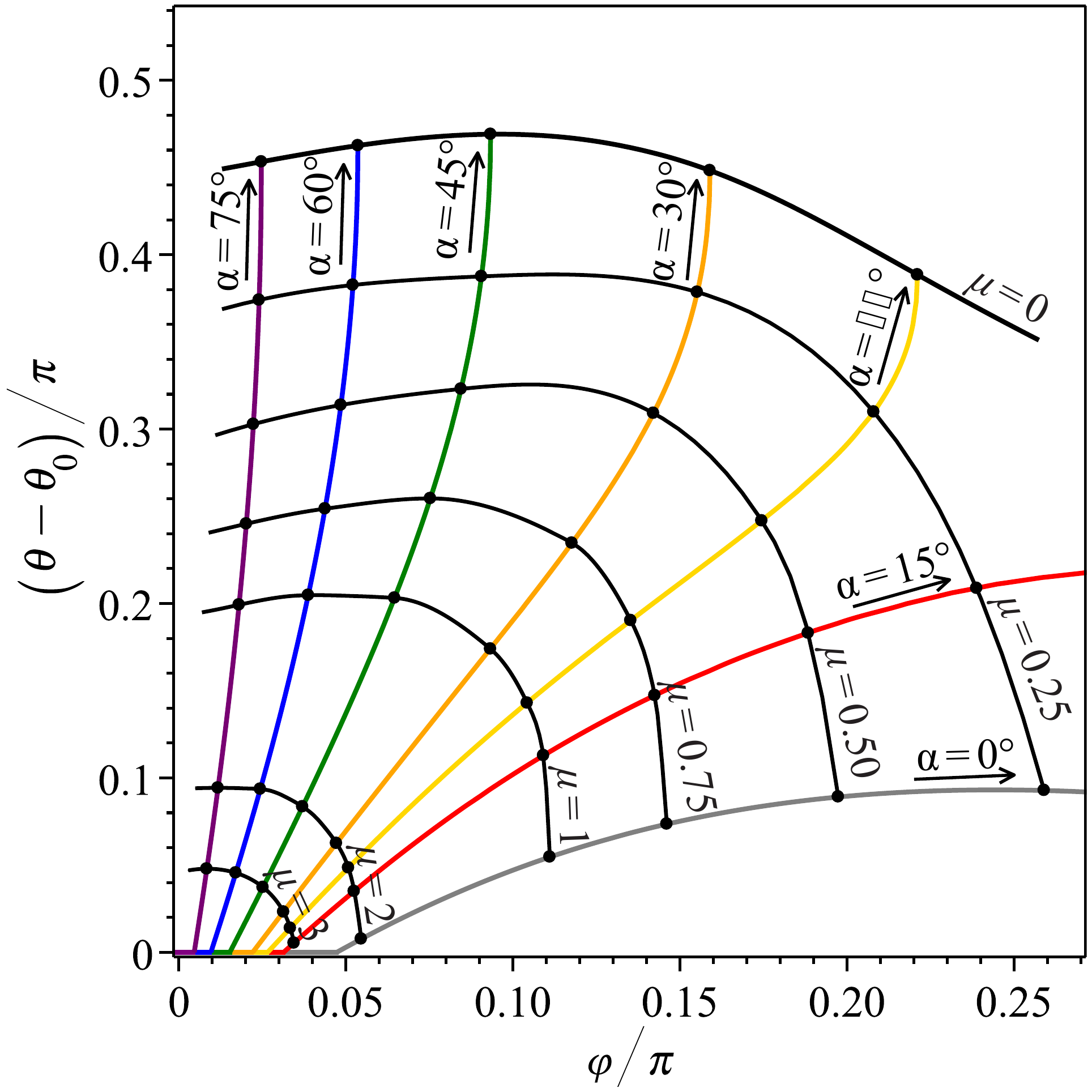}} \\
\caption[]{The plot representation
of the biomimetic scale-covered beam under twisting differentiated to three distinct regimes of
performance including: linear (before scales engagement), kinematically determined nonlinear (during scales engagement), and
a frictional locking boundary for various coefficients of friction: 
\subref{Fig5a} plot of the system for different $\eta$ with the given values of $\theta_0=10^\circ$, $\alpha=45^\circ$, $\beta=1.25$, and $\lambda=0.45$; and
\subref{Fig5b} plot of the system for different $\alpha$ with the given values of $\eta=3$, $\theta_0=10^\circ$, $\beta=1.25$, and $\lambda=0.45$.}%
\label{Fig5}%
\end{figure}

Next, we investigate the scale rotation in response to applied twist. This is achieved by plotting the scale angle rotation $\theta$ versus twist angle $\varphi$. Using nonlinear relationship (\ref{Eq6}), two plots are established spanned by $(\theta-\theta_0)/\pi$ and $\varphi/\pi$ as shown in Figure \ref{Fig5} for different $\eta$ and $\alpha$, respectively.

In Figure \ref{Fig5a}, the given geometrical parameters are as follows $\theta_0=10^\circ$, $\alpha=45^\circ$, $\beta=1.25$, and $\lambda=0.45$. For $\mu = 0$, which indicates frictionless case, we obtain purely kinematic locking points for each $\eta$ by using $\partial \varphi / \partial \theta=0$ to obtain rigidity envelope \cite{c42}. We juxtapose this with plots the rough interfaces ($\mu>0$), where the locking limits are found via the singularity point of friction force described in (\ref{Eq12}). Clearly, friction advances the locking configuration. However, the locking line does not merely translate downwards as observed in the bending case \cite{c39}. This is an important distinction from the pure bending of rough biomimetic beams reported earlier \cite{c39}. As coefficient of friction increases, the frictional locking envelope can intersect the horizontal axis. This is the instantaneous locking or the ``static friction" lock case.

In Figure \ref{Fig5b}, the effect of scales orientation with angle $\alpha$ is investigated. This angle serves as an important geometric tailorability parameter of the system \cite{c42}. In this plot, $\eta=3$, $\theta_0=10^\circ$, $\beta =1.25$, and $\lambda=0.45$. For higher angles $\alpha$, a quicker engagement occurs with steeper nonlinear gains and earlier locking. Interestingly, by decreasing $\alpha$ sufficiently, the system would not reach to the kinematic locking. However, frictional locking is universal and will thus determine the locking behavior. In this aspect, this system again differs from bending case since friction can cause locking even when no-kinematic locking is possible. This figure also shows the possibility of static friction locking for increasing $\mu$. However note that as $\alpha$ increases, such static friction lock becomes more difficult to achieve requiring much higher frictional coefficients. Overall the frictional locking envelope is a highly nonlinear function admitting no closed form solution unlike the pure bending case \cite{c39}.

%%%%%%%%%%%%%%%%%%%% Figure 5 %%%%%%%%%%%%%%%%%%%%
%%%%%%%%%%%%%%%%%%%%%%%%%%%%%%%%%%%%%%%%%%%%%%%%%%

\begin{figure}[htbp]%
\centering
\subfigure[][]{%
\label{Fig6a}%
\includegraphics[width=3in]{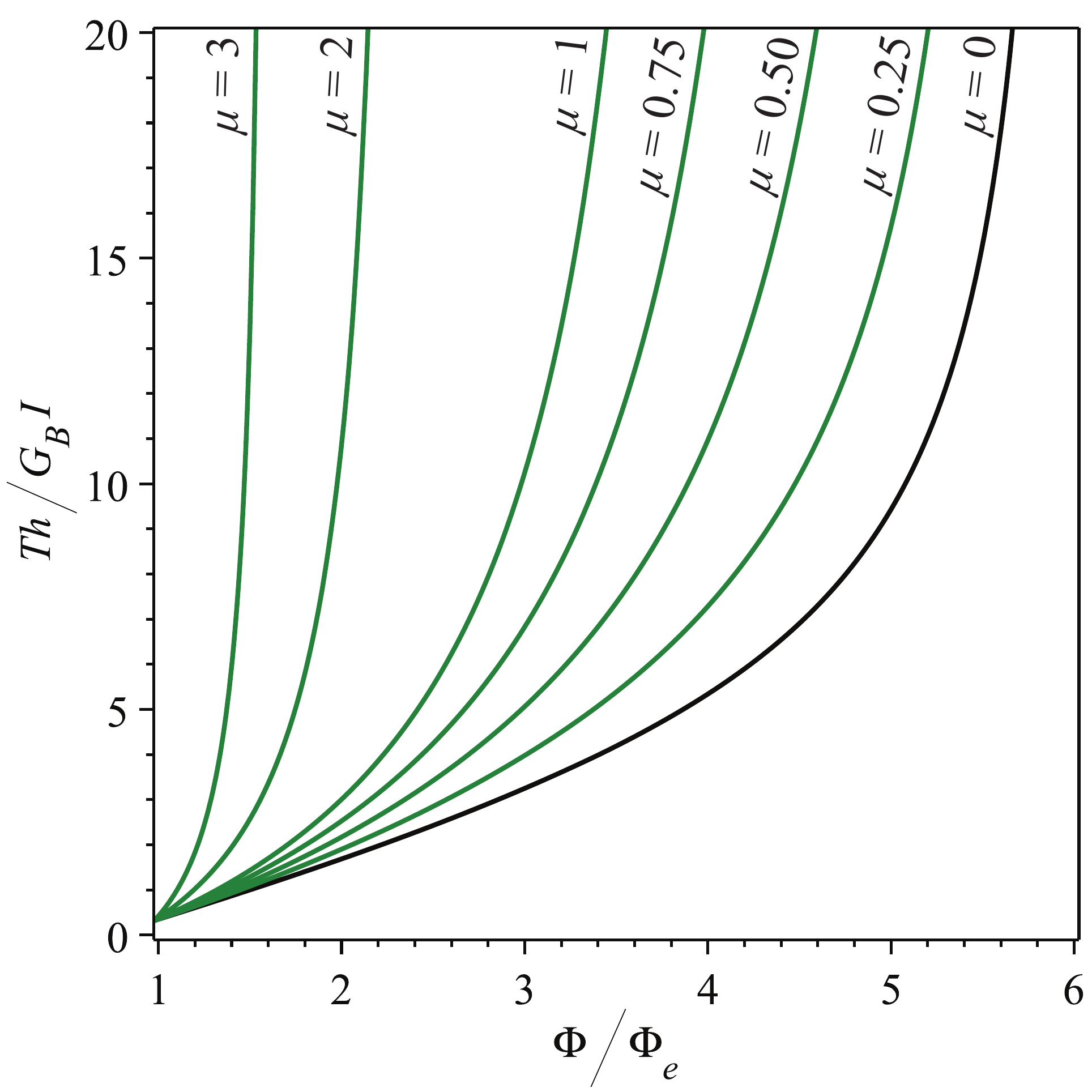}}%
\hspace{8pt}%
\subfigure[][]{%
\label{Fig6b}%
\includegraphics[width=3in]{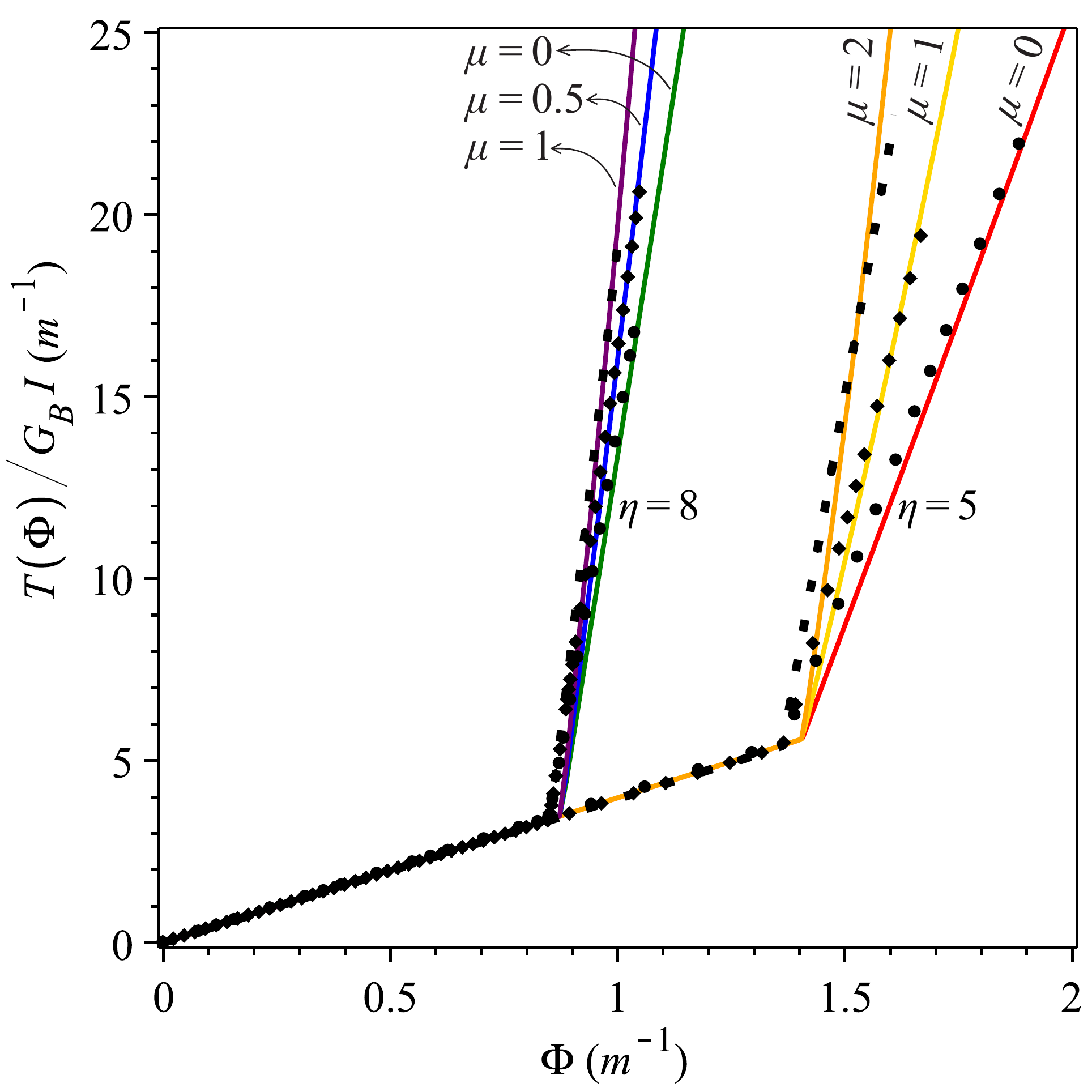}} \\
\caption[]{Torque--twisting rate curve derived from (\ref{Eq14}) for different cases: 
\subref{Fig6a} non-dimensionalized post-engagement torque--twisting rate curves for various coefficients of friction with the given values of $\eta=3$, $\theta_0=10^\circ$, $\alpha=45^\circ$, $\beta=1.25$, $\lambda=0.45$, $\zeta=0.35$, and $L/t_s=35$, showing the perceptible effect of friction in the effective torsional stiffness of the biomimetic scale-covered structure; and
\subref{Fig6b} verification of analytical model using numerical results through the plot of ${T(\Upphi )}/{G_B}I$ versus twist rate ($\Upphi$) for various coefficients of friction and two different $\eta$ with the given values of $\theta_0=10^\circ$, $\alpha=45^\circ$, $\beta=0.6$, $\lambda=0.32$, $\zeta=0.18$, and $L/t_s=45$. Black dotted lines represent FE results.}%
\label{Fig6}%
\end{figure}

In order to understand the effect of friction force on the mechanics of the system, we use (\ref{Eq14}) to plot the non-dimensionalized post-engagement torque–-twisting rate plot for various coefficients of friction, Figure \ref{Fig6a}. Dimensionless geometrical parameters for this case are $\eta=3$, $\theta_0=10^\circ$, $\alpha=45^\circ$, $\beta=1.25$, $\lambda=0.45$, $\zeta=0.35$, and $L/t_s=35$. To verify the analytical model, we have developed an FE model as described in section \ref{Finite element simulations}. Then we have done FE simulations for different $\eta$ and $\mu$ values and extracted torsional response of the structure $T(\Upphi)/G_B I$, versus twist rate from the beginning of the simulation as shown in Figure \ref{Fig6b}. The following dimensionless parameters are used for this model: $\theta_0=10^\circ$, $\alpha=45^\circ$, $\beta=0.6$, $\lambda=0.32$, $\zeta=0.18$, and $L/t_s=45$. Also the following elastic properties are considered for substrate: $E_B=25$ $GPa$, $\nu=0.25$, with a cross section dimension of $32 \times 16$ $mm$. In this figure, the dotted lines are representing FE results. The plot highlights remarkable agreement between analytical and FE results for two different overlap ratios along different coefficients of friction. The small deviation between results could be caused by edge effects and numerical issues.

As shown in Figure \ref{Fig6a}, higher coefficient of friction significantly increases the torsional stiffness of the structure. Therefore, the friction force has a dual contribution to the mechanical response of biomimetic scale-covered system –- while advancing locking, thereby limiting range of motion but also increasing the torsional stiffness of the system.

%%%%%%%%%%%%%%%%%%%% Figure 6 %%%%%%%%%%%%%%%%%%%%
%%%%%%%%%%%%%%%%%%%%%%%%%%%%%%%%%%%%%%%%%%%%%%%%%%

\begin{figure}[htbp]%
\centering
\subfigure[][]{%
\label{Fig7a}%
\includegraphics[width=3in]{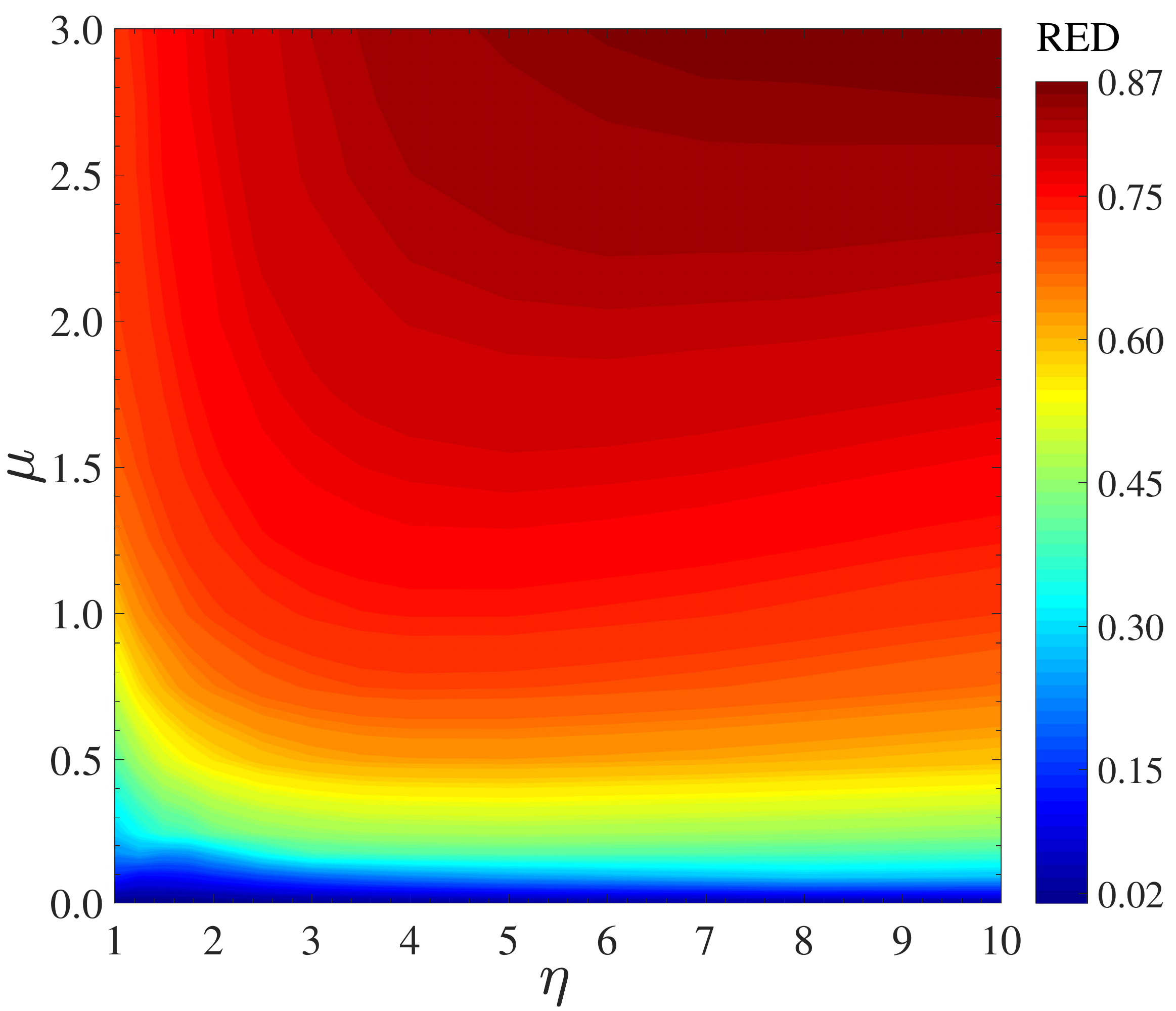}}%
\hspace{8pt}%
\subfigure[][]{%
\label{Fig7b}%
\includegraphics[width=3in]{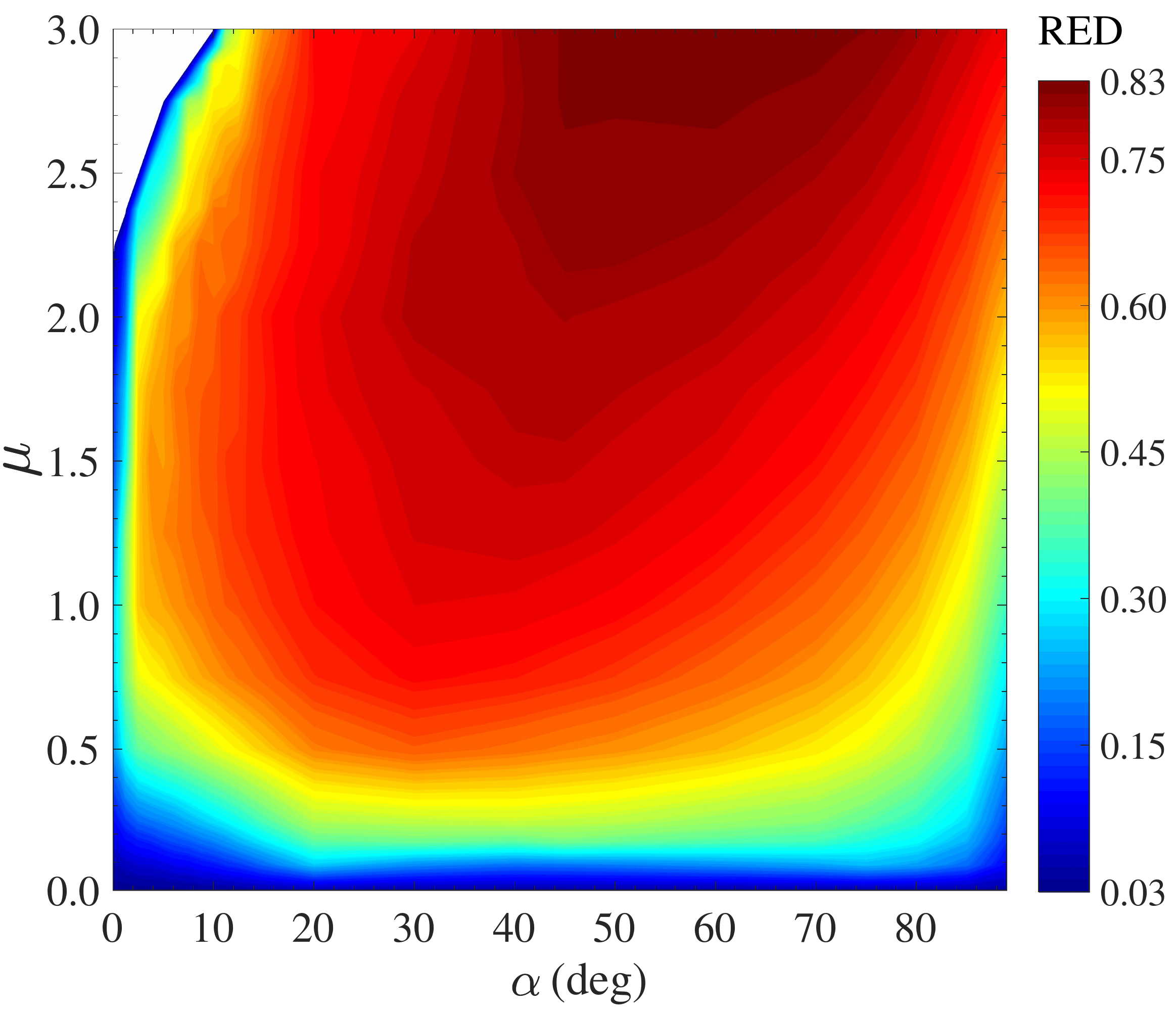}} \\
\caption[]{Non-dimensional relative energy dissipation ($RED$) factor contour plot with given values of $\theta_0=10^\circ$, $\beta=1.25$, $\lambda=0.45$, $\zeta=0.35$, $L/t_s=35$, $E_B=25$ $GPa$, $\nu=0.25$, and the substrate's cross section of $32 \times 16$ $mm$ for two different cases: 
\subref{Fig5a} spanned by $\mu$ and $\eta$ for fixing $\alpha=45^\circ$; and
\subref{Fig5b} spanned by $\mu$ and $\alpha$ for fixing $\eta=3$.}%
\label{Fig7}%
\end{figure}

In order to quantify the dual contribution of friction, we investigate the frictional work during twisting by using the relative energy dissipation ($RED$), described in (\ref{Eq16}). Fixing all parameters involved in $RED$, except $\mu$, $\eta$, and $\alpha$ for the current simulation leads to contour plots shown in Figure \ref{Fig7}. In these contour plots, we have considered $\theta_0=10^\circ$, $\beta=1.25$, $\lambda=0.45$, $\zeta=0.35$, $L/t_s=35$, and the substrate’s properties as follows $E_B=25$ $GPa$, $\nu=0.25$, and the cross section dimension of $32 \times 16$ $mm$.

In Figure \ref{Fig7a}, we fix $\alpha=45^\circ$ to obtain an energy dissipation contour plot spanned by $\eta$ and $\mu$. This plot indicates that $RED$ increases for higher $\mu$, and also increases very slightly with $\eta$. This contour plot shows that $\eta$ does not have as strong effect as coefficient of friction on frictional energy dissipation of the system. However, the frictional work quickly saturates with higher coefficient of friction for all $\eta$. 

To obtain Figure \ref{Fig7b}, we fix $\eta=3$ and the $RED$ contour plot spanned by $\alpha$ and $\mu$. This plot shows that, despite that locking twist rate $\Upphi_{lock}$ increases by decreasing $\alpha$ according to Figure \ref{Fig7b}, the effect of the friction is higher at the range of $40^\circ<\alpha<60^\circ$, and the $RED$ passes through its maximum by increasing $\mu$ around this range of $\alpha$. Also at lower $\alpha$, unilaterally increasing $\mu$ does not necessarily increase the frictional dissipation. The white region in this contour plot is related to the instantaneous post-engagement frictional locking, which happens at lower $\alpha$ and higher $\mu$. At this condition, the system lock statically at the engagement point and the friction force does not work on the system.

%%%%%%%%%%%%%%%%%%%% Figure 7 %%%%%%%%%%%%%%%%%%%%
%%%%%%%%%%%%%%%%%%%%%%%%%%%%%%%%%%%%%%%%%%%%%%%%%%

\section{Conclusion} \label{Conclusion}

We investigate for the first time, the effect of Coulomb friction on the twisting response of a biomimetic beam using a combination of analytical and FE model. We established the extent and limits of universality of frictional behavior across bending and twisting regimes. The analytical model which have been developed, would help in obviating the need for full-scale FE simulations, which are complicated for large number of scales and for large deflection. We find that several aspects of the mechanical behavior show similarity to rough bending case investigated earlier. At the same time, critical differences in response were observed, most notably the effect of the additional dihedral angle. This work shows the dual contribution of frictional forces on the biomimetic scale-covered system, which includes advancing the locking envelope and at the same time adding to the torsional stiffens. Interestingly, if the coefficient of friction is large enough for a given configuration, it can lead to the instantaneous post-engagement frictional locking known as the static friction locking. This investigation demonstrates that engineering scale surfaces to produce wide range of coefficients of friction can play an important role on tailoring the deformation response of biomimetic scales systems under a variety of applications.

%%%%%%%%%% Insert bibliography here %%%%%%%%%%%%%%
\section*{References}
\bibliography{main}

\end{document}